\documentclass{article}

\usepackage{arxiv}

\usepackage{float}
\usepackage{hyperref}

\usepackage{graphicx}
\usepackage{textcomp}
\usepackage{array,ragged2e}
\usepackage{colortbl}
\usepackage{makecell}
\usepackage{pgfplots}
\pgfplotsset{compat=1.17}
\usepackage{amsmath}
\usepackage{multicol}
\usepackage{multirow}
\usepackage{tabto}
\usepackage{wrapfig}
\usepackage{amssymb}
\usepackage{amsmath}

\usepackage{graphicx}
\usepackage[misc,geometry]{ifsym}
\usepackage{lipsum}
\usepackage{fancyvrb}

\usepackage{textcomp}
\usepackage{array,ragged2e}
\usepackage{pgfplots}
\usepackage{url}
\hypersetup{
  linkcolor  = red!60!black,
  citecolor  = blue!90!white,
  urlcolor   = violet!60!black,
  colorlinks = true,
}

\usepackage{orcidlink}

\usepackage{svg}
\usepackage{xcolor}
\newcommand{\orcid}[1]{\href{https://orcid.org/#1}{\includesvg[width=10pt]{orcid.svg}}}

\title{Query Based Construction of Chronic Disease Datasets}

\author{
  Vuong M. Ngo \\
  School of Computing\\
  Dublin City University\\
  Dublin, Ireland \\
  \texttt{vuong.ngo@dcu.ie} \\
   \And
Geetika Sood\\
  School of Computing\\
  Dublin City University\\
  Dublin, Ireland \\
  \texttt{geetika.sood@dcu.ie} \\
     \And
Patricia Kearney\\
  School of Public Health\\
  University College Cork\\
  Cork, Ireland\\
  \texttt{patricia.kearney@ucc.ie} \\
     \And
Fionnuala Donohue\\
    National Health Intelligence Unit\\
    Health Service Executive\\
    Dublin, Ireland \\
  \texttt{fionnuala.donohue2@hse.ie} \\
\And
Dongyun Nie\\
   ADAPT Centre\\
  Dublin City University\\
  Dublin, Ireland \\
  \texttt{dongyun.nie@dcu.ie} \\
\And
 Mark Roantree \\
  Insight Centre for Data Analytics\\
  Dublin City University\\
  Dublin, Ireland \\
  \texttt{mark.roantree@dcu.ie} \\
}

\begin{document}
\maketitle

\begin{abstract}
The RECONNECT project addresses the fragmentation of Ireland's public healthcare systems, aiming to enhance service planning and delivery for chronic disease management. By integrating complex systems within the Health Service Executive (HSE), it prioritizes data privacy while supporting future digital resource integration. The methodology encompasses structural integration through a Federated Database design to maintain system autonomy and privacy, semantic integration using a Record Linkage module to facilitate integration without individual identifiers, and the adoption of the HL7-FHIR framework for high interoperability with the national electronic health record (EHR) and the Integrated Information Service (IIS). This innovative approach features a unique architecture for loosely coupled systems and a robust privacy layer. A demonstration system has been implemented to utilize synthetic data from the Hospital Inpatient Enquiry (HIPE), Chronic Disease Management (CDM), Primary Care Reimbursement Service (PCRS) and Retina Screen systems for healthcare queries.  Overall, RECONNECT aims to provide timely and effective care, enhance clinical decision-making, and empower policymakers with comprehensive population health insights.
\end{abstract}


%
%
%
%
\section{Introduction}
\label{sec:Intro}
In Ireland, chronic diseases like diabetes and cardiovascular conditions place a significant strain on the healthcare system, necessitating coordinated, long-term management. Integrated care models, supported by digital health technologies and national programs, are essential for improving patient outcomes, reducing hospitalizations, and managing healthcare costs effectively \cite{ngo2024hl7}. 
Ireland's healthcare system is under increasing pressure due to a growing and aging population, rising chronic diseases, and evolving patient expectations. These factors have led to longer waiting lists and overcrowded hospitals. The COVID-19 pandemic further highlighted vulnerabilities in infrastructure and staffing although it also provided an opportunity for researchers from different communities to address  issues with dataset engineering \cite{Scriney2023}.  Without intervention, vulnerable populations—such as the elderly and those from lower socioeconomic backgrounds—risk poorer health outcomes. However, this challenge presents an opportunity for transformation through digital technologies and data-driven approaches. The Connecting Government 2030 Strategy promotes digitalization in healthcare, aiming to improve access, enhance efficiency, and position Ireland as a leader in innovative health solutions.

The specific problem to be addressed is the fragmentation and isolation of data within the Irish healthcare system. Currently, health data is scattered across multiple disconnected systems, creating challenges for patients, clinicians, and policymakers. Patients navigating the healthcare service must repeatedly report and recall their health information. Clinicians face difficulties accessing a comprehensive view of a patient's health, compounded by the burden of retrieving information from various sources. For policymakers, making informed decisions about health service delivery is challenging due to the lack of comprehensive information on the population.

Overall, this lack of data integration hampers timely and effective patient care, complicates the delivery of integrated services, and restricts data-driven decision-making in both clinical and strategic contexts. This issue is further exacerbated by the growing pressures on the healthcare system. Inefficient integration and utilization of healthcare data leads to several problems:

\begin{itemize}
    \item Disconnected patient records negatively impact patients, clinicians, and population health planners. These gaps lead to suboptimal patient care, increased workloads for clinicians, and hinder the effectiveness of health planners and policymakers.
    \item Inflexible systems struggle to identify and adapt to changing healthcare needs and priorities.
    \item Manual data processing is not only resource-intensive but also susceptible to errors, further straining an already overwhelmed system.
\end{itemize}

To address these challenges, the proposed solution will focus on creating a comprehensive healthcare data integration infrastructure, which includes:

\begin{itemize}
    \item Structural Integration: A federated database design that preserves the autonomy of existing systems while facilitating varying levels of privacy and access.
    \item Semantic Integration: A record linkage module that complies with data governance policies, allowing for integration even without a universal identifier. This approach can be applied to systems such as HIPE, CDM, PCRS, and RetinaScreen.
    \item Adoption of Standards: A new framework based on the Fast Healthcare Interoperability Resources (HL7-FHIR) model \cite{FHIR2021}, ensuring high levels of interoperability within integrated FHIR data, regardless of the participating healthcare systems.
\end{itemize}

The remainder of this paper is structured as follows.
In Section \S\ref{sec:method}, we outline the four-layer architecture used in creating digital healthcare assets. Section
\S\ref{sec:construction} details the generation of healthcare assets, including our integration strategy used with examples.
Finally, in Section \S\ref{sec:conc}, we present our conclusions.

\section{Methodology}
\label{sec:method}

\begin{figure}[h]
\centering
\includegraphics[scale = 0.5]{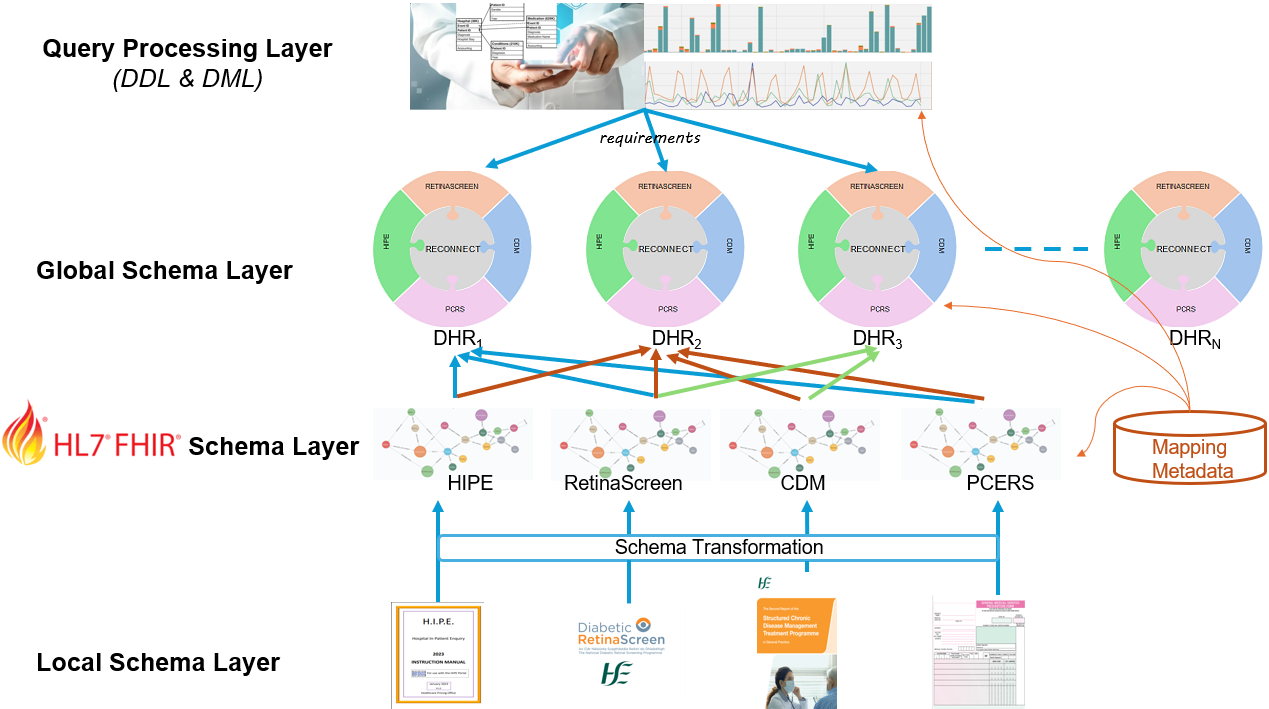}
\caption{RECONNECT Architecture: Creation and Manipulation of Novel Digital Healthcare Assets}
\label{fig:Arch_overall}
\end{figure}

In this section, we outline the RECONNECT methodology for creating digital healthcare assets, specifically datasets based on either CSV or FHIR formats. Our approach utilizes a four-layer \emph{federated} architecture that encompasses data at various levels of structure and representation. As illustrated in Figure \ref{fig:Arch_overall}, a federated architecture \cite{sheth1990federated} operates under the premise that systems are loosely coupled—meaning they do not need to communicate directly with one another—and are read-only, indicating that the RECONNECT architecture cannot write to the systems residing at the Local Schema Layer. This architecture offers a degree of autonomy that, among other benefits, supports a high level of privacy and complies with stringent governance procedures within organizations \cite{hermann2019}.

\subsection{Local Schema Layer}
The Local Schema layer contains all source systems which may be entirely heterogeneous containing systems based on relational databases, hierarchical databases, Web (HTML or XML) sources or proprietary data. This \emph{autonomy} of source information systems is preserved in a federated database architecture which is a crucial feature of RECONNECT: no participating system requires modification nor is required to facilitate updates from with the RECONNECT system.

The Reconnect Local Schema layer replicates four distinct datasets within the HSE that are specific to a particular chronic disease: Type 2 Diabetes. Currently, these systems are fragmented, preventing any digital interoperability. The datasets are as follows:
\begin{enumerate}
    \item	HIPE: Hospital In-Patient Enquiry
    \item 	CDM: Chronic Disease Management
    \item 	PCRS: Primary Care Reimbursement Service
    \item 	Retina Screen
\end{enumerate}

\textbf{HIPE:} The HIPE system maintains the national database of hospital discharge activity, serving the data needs of various stakeholders, including policymakers, clinical teams, and researchers. It is the primary source of national data on hospital discharges, encompassing all acute public hospitals, though private hospitals are excluded from the database. HIPE collects demographic, clinical, and administrative information on discharges and deaths in public hospitals across the country. This dataset also feeds into national files that are used for activity-based funding.

The synthetic dataset is derived from HIPE metadata and represents a scaled-down version of records for the entire population of Ireland. It replicates the statistical properties and patterns of real-world healthcare data without containing any actual identifiable personal information. Each patient is identified by an MRN number, which is encrypted in the national file. The dataset includes limited personal details, as well as information on diagnoses, treatments, and the doctors assigned to each patient discharged from the hospital. Synthetic data has shown to offer significiant advantages when prototyping these types of applications \cite{DC2018}, \cite{CCP2019} in advance of the receipt of actual healthcare datasets. 

\textbf{CDM:} The program is designed for individuals aged 18 and above with specific chronic conditions such as type 2 diabetes, asthma, COPD, and cardiovascular diseases. It includes regular reviews, personalized care plans, medication assessments, support for condition management, early detection of new conditions and complications, and community-based care. The synthetic data is generated based on the structure and guidelines of the CDM treatment program.

\textbf{PCRS:} It is a division of the HSE responsible for reimbursing healthcare professionals for services rendered to the public. PCRS manages payments for high-tech drugs, reimburses hospitals for medications, and oversees the assessment of Medical Card and GP Visit Card applications. Additionally, PCRS compiles statistics and conducts trend analyses for stakeholders, aiding in policy development and strategic decision-making.

\textbf{Retina Screen:} This program provides national diabetic retinopathy screening for individuals aged 12 and above with diabetes. The primary goal of the database is to identify diabetes cases by HSE area, initially to support Retina Screen program, with potential expansion to other aspects of diabetes care. Retina Screen is a population-based database designed to identify and collect data on instances of Type 2 diabetes within a specified population. However, it does not capture all cases of Type 2 diabetes.

A separate PII database is used to store all personally identifiable information (PII). Access to this data is controlled through a role-based security layer, and while the PII dataset is currently used for data integration, it remains separate from the main screening dataset..

\subsection{FHIR Schema Layer}
While systems may have heterogeneous data models at the Local Schema Layer, all datasets which have been extracted form the underlying systems must have a common data representation \cite{batini1986comparative}. Thus, at this stage, data will be extracted from source systems and transformed into the RECONNECT common data model (HL7-FHIR). Additionally, systems will be interconnected using established integration techniques. FHIR has the latest HL7 \cite{saripalle2019using} healthcare standard and has been used in similar projects \cite{franz2015applying}, \cite{walinjkar2018fhir} to varying degrees of success. Similarly graph-based common models such as those used in \cite{Park2014} have been shown to have benefits over the more traditional relational model which is often used in these types of architectures. The adoption of HL7-FHIR offers a high degree of extensibility to the solution presented here.

\textbf{Mapping Metadata:} The purpose of the FHIR mapper is to ensure consistent interpretation of data across systems which is challenging due to variations in coding systems and clinical terminologies. Each data source contains various attributes across different categories. For this, we build upon earlier work on metadata mapping from multiple sources \ref{Scriney2019} as it plays a crucial role in connecting source datasets with their \emph{roles} in systems such as RECONNECT. In FHIR, these categories correspond to distinct resources. Different attribute sets within a data source are mapped to different FHIR resources. Transforming a data source into a FHIR subgraph is achieved using a resource map and a namespace linker.

This study includes four potential mapping types:
\begin{enumerate}
    \item ONE\_TO\_ONE mappings: Applied when there is a direct correspondence between source and FHIR properties, allowing the attribute value from the data source to be directly imported.
    \item MANY\_TO\_ONE mappings: Used when multiple source attributes are needed to populate a single FHIR property.
    \item INDIRECT mappings: Utilized to provide default values for FHIR resources that are absent in the source data.
    \item LOOKUP mappings: Indicate attributes that require record linkage to populate the corresponding FHIR property.
\end{enumerate}

\textbf{Record Linkage:} This step will provide a holistic patient record by linking the databases. Record linkage \cite{nie2019detecting}, a well-known challenge in data integration, is often simple when databases share common identifiers. However, healthcare systems rarely align in terms of structure or identifiers, making it difficult to accurately identify patients and achieve proper integration. In this project, the healthcare systems lacked a single unique identifier, contained unrecognizable or missing identifiers, and recorded patient data inconsistently.

Record linkage typically relies on probabilistic, inexact attribute matching between systems (e.g., name, date of birth and contact details), but these attributes are not available here due to privacy concerns \ref{brizan2006survey}. Previous research has tackled this issue, even within healthcare. In the current architecture, only researchers working with synthetic data can perform record linkage, as shown in Figure \ref{fig:Arch_overall}. They create a Patient Meta-Record, which links unique identifiers from the evaluated systems. This anonymized Meta-Record facilitates data integration from systems assessed for linkage. This approach will address the gap caused by the lack of an Individual Health Identifier (IHI).

\subsection{Global Schema Layer}
The concept is to create a new (distributed) digital asset for each user requirement. These assets may be shared and reused across multiple requirements (with appropriate governance) or utilized for a single case study. They remain in the system as pre-computed queries until the user chooses to delete them and can be updated as needed. The assets we create and populate are referred to as Digital Health Records (DHRs).

A global schema provides a standardized digital data asset for capturing and presenting patient information or offering an overview of the prevalence of chronic diseases and associated risk factors across various demographics, age groups, and genders. This helps identify patients with similar risk factors and predict the likelihood of chronic disease development in individuals. Section \ref{sec:constructiontech} outlines how this new system can be adapted to generate multiple distributed digital assets based on different user requirements. These digital assets can be shared among multiple users, with access determined by the level of permissions granted.

Digital health records are invaluable resources for both health research and clinical decision-making. They provide a comprehensive view of patient history, treatments, and outcomes, which can be used to enhance patient care and support public health initiatives.

\begin{figure}[h]
\centering
\includegraphics[scale = 0.6]{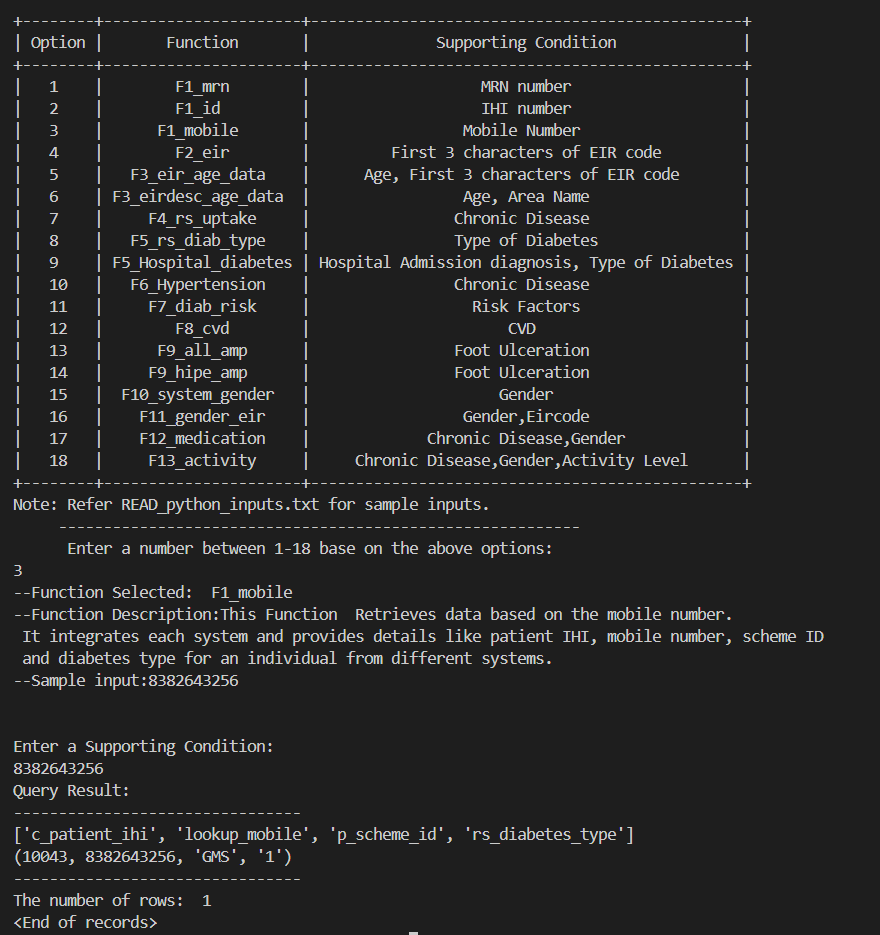}
\caption{System interaction with a single parameter}
\label{fig:Sysoutput1}
\end{figure}

\begin{figure}[h]
\centering
\includegraphics[scale = 0.52]{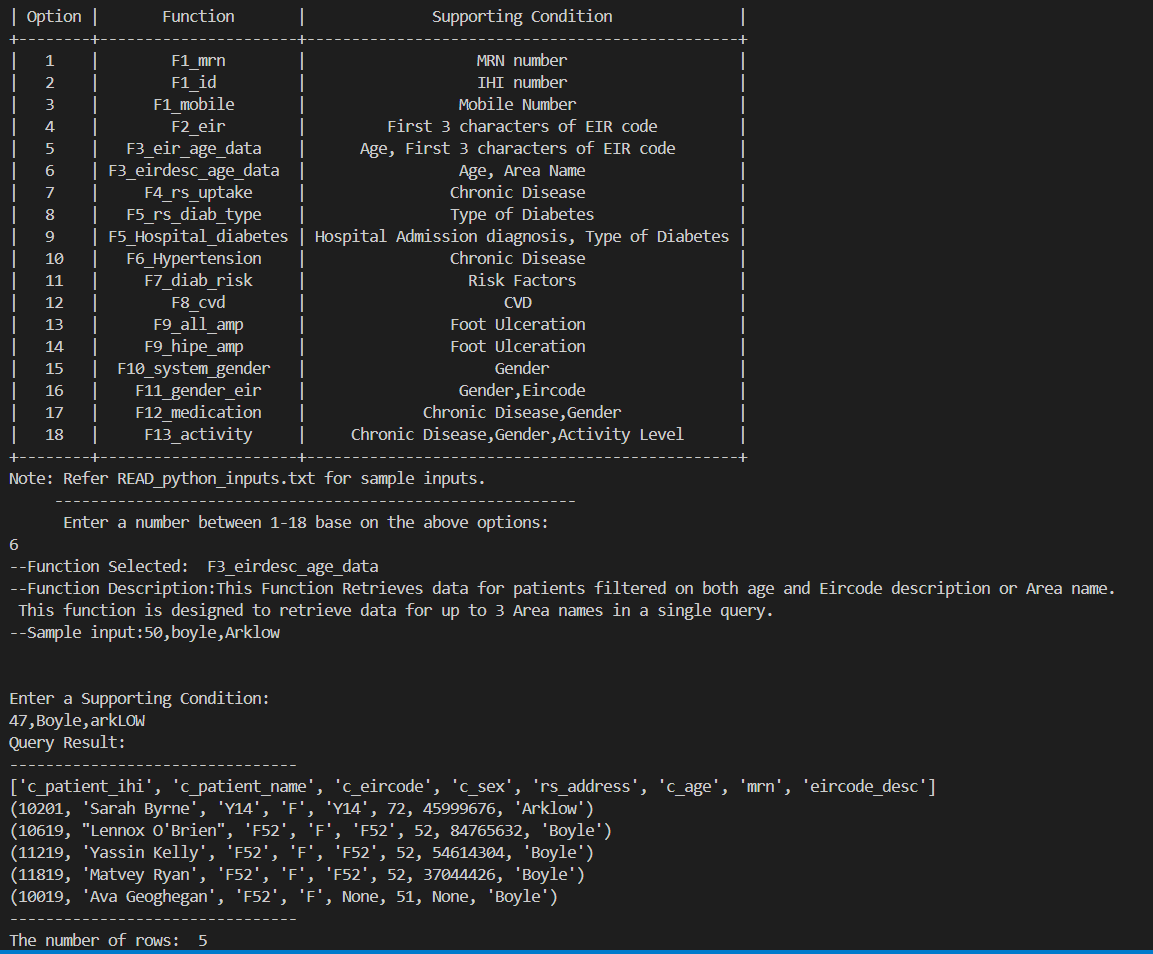}
\caption{System interaction with multiple parameters}
\label{fig:Sysoutput2}
\end{figure}

Figures \ref{fig:Sysoutput1} and \ref{fig:Sysoutput2} illustrate interactions with the system, where users select a number from the provided list and input the relevant "where" clause to obtain results. This process generates the expected output for case  \ref{F1}, with the digital health record dynamically populated in the global schema layer using data from the HIPE, CDM, PCRS, and RetinaScreen systems. Similar results can be achieved by selecting other functions and entering a corresponding "where" clause, as demonstrated in the samples provided in Section \ref{sec:constructiontech}

Comprehensive interoperability can be realized by integrating national systems to improve patient care, ensure regulatory compliance, enhance operational efficiency, extend the reach of national prevention programs, and optimize cost management within healthcare organizations.

\subsection{Query Processing Layer}

There are 3 "case studies" in the illustration: blue (pulling from 3 sources); brown (pulling from 4 sources); green (2 sources).
\begin{enumerate}
    \item Uptake of Retina Screen among People with Diabetes. This case study refers to the percentage of people not participating in the prevention services led by the government but ending up in the hospital. Datasets used:
\begin{enumerate}
    \item Retinascreen
    \item CDM
    \item HIPE
\end{enumerate}

Diabetic retinopathy (DR) is the leading cause of preventable blindness. The independent risk factors for DR included diabetes duration, haemoglobin A1c, serum glucose, systolic blood pressure, and duration of diabetes. After 5 years, approximately 25\% of type 1 diabetes patients will have retinopathy. After 10 years, almost 60\% will have retinopathy, and after 15 years, 80\% will have retinopathy. International guidelines for diabetic retinopathy (DR) screening, released by the International Council of Ophthalmology (ICO), specify that adequate DR screening should encompass a visual acuity test and a retinal examination. 

This case study aims to pull data from Retinascreen, CDM and HIPE databases. The data governance layer specifies the level of detail accessible by the system operator.

\item Blood Pressure Control among People with Diabetes

Datasets: 
\begin{enumerate}
    \item PCRS
    \item CDM
    \item HIPE
\end{enumerate}

Randomized controlled trials have shown that lowering systolic blood pressure (SBP) to less than 140 mmHg and diastolic blood pressure (DBP) to less than 90 mmHg benefits people with diabetes. If SBP is 140 mmHg or more and/or DBP is 90 mmHg or more, drug therapy is necessary, preferably starting with a combination therapy. The use of renin-angiotensin system (RAS) inhibitors is strongly supported, especially in patients with evidence of end-organ damage. Controlling blood pressure often requires multiple drug therapies, and a combination of two or more drugs at fixed doses in a single pill should be considered to improve adherence and achieve earlier control of blood pressure.

\item Amputations among People with Diabetes

Datasets:
\begin{enumerate}
    \item CDM
    \item HIPE
\end{enumerate}

Diabetes can lead to foot or leg amputation, with a limb amputated every 20 seconds globally due to diabetes. 85\% of these amputations are preceded by a foot ulcer. The HSE introduced the National Diabetes Footcare program in 2010, recommending annual foot screenings for people with diabetes to assess their risk of lower extremity amputation. Those at risk should be referred to foot protection services in the community or hospital setting.

\end{enumerate}

\section{Generating Healthcare Assets}
\label{sec:construction}

 The healthcare assets represent a subset of the primary integrated datasets, specifically designed to meet the needs of the system’s end-users. These assets are tailored to serve both individual users, such as clinicians, and larger teams like the health intelligence unit. By providing read-only access, they effectively prevent accidental data modifications. Built on integrated data, they enable faster and more efficient workflows. Access to these assets is controlled through role-based security to ensure that only authorized users can retrieve the data.

These healthcare assets are sourced from existing datasets, using dynamic queries that add an extra layer of data security. Depending on the use case and access permissions, end-users can retrieve integrated data, including Personally Identifiable Information (PII). Various healthcare assets are available for analysis, focusing on specific conditions of national and local interest, such as Type 2 diabetes and related conditions like hypertension, retinopathy, and amputations.

The Individual Health Identifier (IHI) is crucial to the data integration process, enabling the identification of individual patients within the dataset. The IHI National Register has been established with 4,775,629 records sourced from a recognized data reservoir. Each record in this repository is assigned a unique IHI number, formally initiating the IHI National Register. In cases where IHI is unavailable, a separate key is generated using fuzzy matching techniques, based on attributes identified across different systems.

\begin{figure}[h]
\centering
\includegraphics[scale = 0.9]{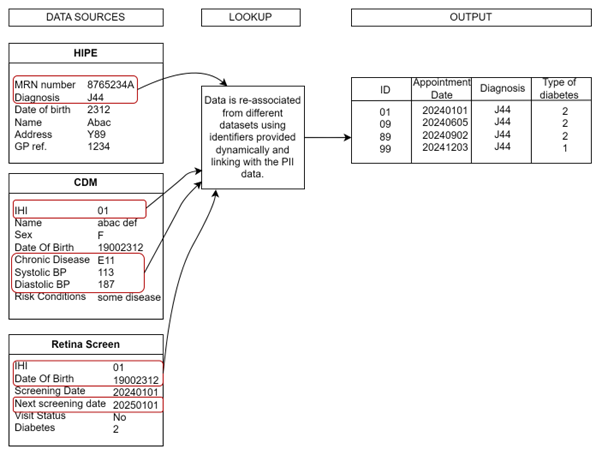}
\caption{High-Level example of how Diabetic patients who missed their Retinopathy appointment are identified}
\label{fig:High-Level-Example}
\end{figure}

Different Healthcare assets developed include:
\begin{enumerate}
 \item 	Integrating HIPE, CDM, PCRS, and Retinascreen systems for an Individual: This asset involves the comprehensive integration of health data from HIPE, CDM, PCRS, and RetinaScreen for individual patients. The goal is to provide a holistic view of a patient’s health status and enhance care coordination, as shown in Figure \ref{fig:High-Level-Example}.
 
\item 	Uptake of Retina Screen among People with Diabetes: Since retinopathy affects individuals with both Type 1 and Type 2 diabetes, this dataset tracks the number of people who have undergone RetinaScreen based on their type of diabetes. Understanding these numbers can help identify gaps in care and improve screening practices.

\item  Uptake of Retina Screen among People on Type of Diabetes: Since retinopathy affects individuals with both Type 1 and Type 2 diabetes, this dataset tracks the number of people who have undergone RetinaScreen, categorized by their type of diabetes. Analyzing these numbers can help identify gaps in care and improve screening practices.

\item 	Multimorbidity (Prevalence of More Than One Chronic Disease): This dataset focuses on the prevalence and management of individuals with multiple chronic diseases, also known as multimorbidity. It highlights the need for comprehensive care strategies to address the complexities of managing multiple health conditions simultaneously. The CDM system tracks individuals diagnosed with Type 2 diabetes, asthma, chronic obstructive pulmonary disease (COPD), and cardiovascular diseases, including stable heart failure, ischaemic heart disease, cerebrovascular disease (stroke/TIA), and atrial fibrillation. This dataset helps identify subsets of individuals with similar underlying risk factors.

\item 	Diabetes, Medication, and Physical Activity: This dataset explores the relationship between diabetes management, medication usage, and the role of physical activity. It emphasizes the impact of lifestyle changes, particularly exercise, on medication requirements and overall diabetes control. Engaging in physical activity fewer than three times per week is identified as a risk factor for developing chronic diseases.

\item 	Age, Physical Activity, and Hospital Admission due to Chronic Disease: This dataset explores the correlation between a patient’s age, level of physical activity, and the frequency of hospital admissions related to chronic diseases. It emphasizes the importance of promoting physical activity, particularly among older adults, to reduce hospitalizations.

\item 	Blood Pressure Control among People with Diabetes: This dataset focuses on the critical need to manage blood pressure in individuals with diabetes. Drug therapy is recommended for those with diastolic blood pressure above 90 and systolic blood pressure above 140. Proper blood pressure control is essential for preventing complications such as stroke, coronary events, and kidney disease.

\item 	Cardiovascular Disease among People with Diabetes: This dataset identifies diabetes patients who are at an increased risk of developing cardiovascular disease. It highlights the importance of regular screenings and preventive measures to mitigate this risk.

\item	Hospital Admissions among People with Diabetes: This dataset compiles data from patients registered in both the CDM and HIPE systems, focusing on the number of hospital admissions for individuals diagnosed with Type 2 diabetes. Its goal is to identify patterns and causes of hospitalizations, improving diabetes management and reducing healthcare costs.

\item	Amputations among People with Diabetes: This dataset tracks the incidence of amputations in individuals with diabetes, often resulting from complications like neuropathy, poor circulation, and other risk factors. It highlights the importance of preventive care, such as regular foot screenings and early interventions, to reduce diabetes-related amputations. 

\item 	Identifying Patients in a Demographic Location Based on Gender: This dataset categorizes patients within specific demographic locations by gender. The data can be used to tailor healthcare services and design targeted outreach programs.

\item 	Identifying Patients in a Demographic Location Based on Age: This dataset focuses on identifying patients within specific demographic locations, categorized by age. Understanding the age distribution allows healthcare providers to address the specific needs of different age groups more effectively. Individuals aged 45 and older are more prone to developing chronic diseases and other risk factors.

\item 	Medications for People with Both Diabetes and Hypertension: This dataset examines the various medications prescribed to individuals managing both diabetes and hypertension. It underscores the importance of addressing both conditions concurrently to reduce health risks. This data also aids clinicians in developing better care models for these patients

\item 	Identify Patient Subgroups with Shared Conditions: This dataset identifies subgroups of patients with similar health conditions, enabling healthcare providers to develop targeted interventions. Such insights can improve the effectiveness of treatment plans and enhance patient outcomes.

\end{enumerate}

\subsection{Technical Methodology}
\label{sec:constructiontech}
In this section, we provide some detail on the integration strategy adopted for the construction of the healthcare assets described previously.

The dynamic\_case\_procedure is a PostgreSQL stored procedure for dynamic data retrieval. It allows users to retrieve specific health-related data from the local schema layer, based on a given condition\_type and an identification parameter. The procedure creates a temporary table named result\_value to store the results returned by the called functions. It calls various functions based on the value of condition\_type.

\textbf{Explanation for Dynamic procedure.}
\begin{enumerate}
    \item Parameters:

condition\_type: Used to determine which query to run based on the provided value.

identification: An identifier (could be MRN, ID, mobile, etc.) to be used in the SQL queries.
\item Dynamic Table Creation:
The procedure dynamically creates a table named result\_value based on the type of data being fetched, which is determined by the condition\_type provided.
\item CASE Structure:
Depending on the value of condition\_type, a corresponding query is executed.
Each case uses a specific function (e.g., hipe\_data, cdm\_data, etc.) to retrieve data from various tables. 

\item ELSE Clause:
If none of the provided conditions match, the procedure raises a notice saying "Check selected function"
\end{enumerate}

This procedure is designed to dynamically generate queries and create tables based on the input condition\_type and identification. 

The procedure calls a function internally. Each function is designed to retrieve patient-specific health records by combining data from multiple tables using dynamic SQL. In the case where no matching condition\_type is found, the procedure raises a notice indicating that no matching record was found. This provides feedback to the user or calls the application without terminating the procedure abruptly. The procedure can be called running a Python script and providing the related parameters. Each function is based on the following:
\begin{enumerate}
    \item Parameters: The function accepts several parameters, including the table names (TABLE1, TABLE2, TABLE3, TABLE4, TABLE5, TABLE6) and filtering conditions (WHERECLAUSE1, WHERECLAUSE2, WHERECLAUSE3).
    \item Dynamic SQL Query: The query variable is constructed using the format function, which dynamically inserts table names and WHERE conditions into the SQL statement.
    \item Joins: It performs several JOIN operations between the tables to gather patient data like MRN, IHI, contact details, diagnosis, and\/or screening details.
    \item WHERE Clause: The query filters data based on screening date, diagnosis, and chronic diseases.
    \item Execution: The dynamically generated query is executed, and the result set is returned using RETURN QUERY EXECUTE query.
\end{enumerate}

\textbf{Procedure Calls and Functionality.} 

The procedure performs the following actions based on the condition\_type and displays results in a dynamically created result table. Sample query follows every case. The following queries contain the condition\_type used:

\begin{enumerate}
\item	F1\_mrn\label{F1}: Example \ref{ex:query1} retrieves data based on the mrn (medical record number) using the hipe\_data function. This function integrates all the systems in the local schema and provides details regarding each individual patient from all systems.

\newenvironment{example}{\hfill\break\begin{center}}{\end{center}}

\begin{example}
\begin{BVerbatim}
call dynamic_case_procedure('F1_mrn','10164260');
select * from result_value;
\end{BVerbatim}

\hfill\break
Sample Query 1.

\label{ex:query1}
\end{example}


\item	F1\_id: Example \ref{ex:query2} retrieves data based on the Individual Health Identifiers using the cdm\_data function. This function integrates CDM, PCRS and RetinaScreen systems from the local schema and provides details regarding an individual from all systems.

\begin{example}
\begin{BVerbatim}
call dynamic_case_procedure('F1_id','10043');
select * from result_value;
\end{BVerbatim}

\hfill\break
Sample Query 2.

\label{ex:query2}
\end{example}


\item	F1\_mobile: Example \ref{ex:query3} retrieves data based on the mobile number using the rs\_data function. It provides a similar functionality as the previous function however the mobile number uniquely identifies the patient.

\begin{example}
\begin{BVerbatim}
call dynamic_case_procedure('F1_mobile','8382643256');
select * from result_value;
\end{BVerbatim}

\hfill\break
Sample Query 3.

\label{ex:query3}
\end{example}


\item	F2\_eir: Example \ref{ex:query4} retrieves patient data based on the EIR code using the eir\_data function. The first three characters of the Eircode that identify the area are stored in the database.

\begin{example}
\begin{BVerbatim}
call dynamic_case_procedure('F2_eir','F52');
select * from result_value;
\end{BVerbatim}

\hfill\break
Sample Query 4.

\label{ex:query4}
\end{example}


\item 	F3\_eir\_age\_data: Example \ref{ex:query5} retrieves data for patients filtered on both age and Eircode using the eir\_age function. It takes a minimum age and the first three characters of an Eircode as parameters and retrieves patient details such as name, sex, address, age, and Eircode. This function is designed to retrieve data for up to 3 Eircodes in a single query.

\begin{example}
\begin{BVerbatim}
call dynamic_case_procedure('F3_eir_above45_data','F52');
select * from result_value;
\end{BVerbatim}

\hfill\break
Sample Query 5.

\label{ex:query5}
\end{example}


\item	F3\_eirdesc\_age\_data: Example \ref{ex:query6} retrieves data for patients filtered on both age and Eircode description using the eirdesc\_age\_data function. It takes a minimum age and the area name as parameters and retrieves patient details such as name, sex, address, age, and Eircode description. This function is designed to retrieve data for up to 3 Area names in a single query.

\begin{example}
\begin{BVerbatim}
call dynamic_case_procedure('F3_eirdesc_above45_data','Boyle');
select * from result_value;
\end{BVerbatim}

\hfill\break
Sample Query 6.

\label{ex:query6}
\end{example}


\item  F4\_rs\_uptake: Example \ref{ex:query7} retrieves data related to patients who choose not to enrol in the Retinopathy programme for the prevention of Retinopathy but were admitted to the hospital and diagnosed with Retinopathy.

\begin{example}
\begin{BVerbatim}
call dynamic_case_procedure('F4_rs_uptake','Type 2 diabetes');
select * from result_value;
\end{BVerbatim}

\hfill\break
Sample Query 7.

\label{ex:query7}
\end{example}


\item F5\_rs\_diab\_type: Example \ref{ex:query8} retrieves data related to the patient suffering from diabetes but a distinction on the type of diabetes is made using the rs\_diab\_type function.

\begin{example}
\begin{BVerbatim}
call dynamic_case_procedure('F5_rs_diab_type','1');
select * from result_value;
\end{BVerbatim}

\hfill\break
Sample Query 8.

\label{ex:query8}
\end{example}


\item F5\_Hospital\_diabetes: Example \ref{ex:query9} retrieves data based on hospitalization due to any condition and type of diabetes.

\begin{example}
\begin{BVerbatim}
call dynamic_case_procedure('F5_rs_diab_type','2');
select * from result_value;
\end{BVerbatim}

\hfill\break
Sample Query 9.

\label{ex:query9}
\end{example}


\item F6\_Hypertension: Example \ref{ex:query10} retrieves data related to patients who are diagnosed with both diabetes and hypertension using the diab\_hyp function.

\begin{example}
\begin{BVerbatim}
call dynamic_case_procedure('F6_Hypertension','Type 2 diabetes');
select * from result_value;
\end{BVerbatim}

\hfill\break
Sample Query 10.

\label{ex:query10}
\end{example}


\item 	F7\_diab\_risk: Example \ref{ex:query11} retrieves data related to diabetes risk factors like physical activity, age, chronic diseases or any other risk factors using the diab\_risk function. Other risk factors that have been included are: overweight or obesity, age 45 or older, parent or sibling with type 2 diabetes, being physically active less than 3 times a week, have non-alcoholic fatty liver disease (NAFLD).

\begin{example}
\begin{BVerbatim}
call dynamic_case_procedure('F7_diab_risk', 'sibling with type 2 diabetes, 
                            non-alcholic fatty liver disease,
                            parent with type 2 diabetes, ethnicity, overweight');
select * from result_value;
\end{BVerbatim}

\hfill\break
Sample Query 11.

\label{ex:query11}
\end{example}


\item 	F8\_cvd: Example \ref{ex:query12} retrieves data related to diabetic patients who are also diagnosed with one or more cardiovascular diseases using the diab\_cvd function. Different Cardiovascular diseases mentioned in the CDM booklet are Stable Heart Failure, Ischaemic Heart Disease, Cerebrovascular Disease (Stroke / TIA) and/or Atrial Fibrillation.
\newpage

\begin{example}
\begin{BVerbatim}
call dynamic_case_procedure('F7_diab_risk', 'sibling with type 2 diabetes, ethnicity');
select * from result_value;
\end{BVerbatim}

\hfill\break
Sample Query 12.

\label{ex:query12}
\end{example}


\item 	F9\_all\_amp: Example \ref{ex:query13} retrieves data related to diabetic amputations using the amputation function.

\begin{example}
\begin{BVerbatim}
call dynamic_case_procedure('F8_cvd', 'Atrial fibrillation,Ischaemic Heart Disease,
                          Stroke,Stable Heart Failure');
select * from result_value;
\end{BVerbatim}

\hfill\break
Sample Query 13.

\label{ex:query13}
\end{example}


\item 	F9\_hipe\_amp: Example \ref{ex:query14} retrieves data related to diabetic patients who have had amputations and are registered in the hospital system using the amputation\_hipe function.

\begin{example}
\begin{BVerbatim}
call dynamic_case_procedure('F8_cvd', 'Stable Heart Failure');
select * from result_value;
\end{BVerbatim}

\hfill\break
Sample Query 14.

\label{ex:query14}
\end{example}


\item	F10\_system\_gender: Example \ref{ex:query15} retrieves data based on gender (used as an identifier here) using the gender\_data function.

\begin{example}
\begin{BVerbatim}
call dynamic_case_procedure('F9_all_amp', 'Foot Ulceration');
select * from result_value;
\end{BVerbatim}

\hfill\break
Sample Query 15.

\label{ex:query15}
\end{example}


\item	F11\_gender\_eir: Example \ref{ex:query16} retrieves gender-related data based on Eircode using the gender\_eir\_data function.

\begin{example}
\begin{BVerbatim}
call dynamic_case_procedure('F9_hipe_amp', 'Foot Ulceration');
select * from result_value;
\end{BVerbatim}

\hfill\break
Sample Query 16.

\label{ex:query16}
\end{example}


\item 	F12\_medication: Example \ref{ex:query17} retrieves data regarding the diabetic and hypertension patients’ medication data using the diab\_hyp\_med function to understand treatment provided in different parts of the country.

\begin{example}
\begin{BVerbatim}
call dynamic_case_procedure('F10_system_gender', 'F');
select * from result_value;
\end{BVerbatim}

\hfill\break
Sample Query 17.

\label{ex:query17}
\end{example}


\item	F13\_activity: Example \ref{ex:query18} retrieves activity-related data using the diab\_hyp\_act function. The data can be retrieved using different chronic diseases, gender and physical activity frequency.

\begin{example}
\begin{BVerbatim}
call dynamic_case_procedure('F11_gender_eir', 'F,F93');
select * from result_value;
\end{BVerbatim}

\hfill\break
Sample Query 18.

\label{ex:query18}
\end{example}




\end{enumerate}

\section{Conclusions}
\label{sec:conc}

In this paper, we proposed a method for integrating chronic disease systems to address shortcomings in existing healthcare systems. The RECONNECT system features a generic architecture comprising a Record Linkage component, tailored for environments with loosely coupled information systems. This enhances integration by enabling seamless connections between disparate data sources. Unlike traditional systems, RECONNECT adheres to the global HL7-FHIR standard, improving interoperability and aligning with international best practices. Additionally, it efficiently reuses digital assets, enhancing healthcare delivery. Privacy and data security are prioritized, with a dedicated privacy layer safeguarding patient information. A prototype using synthetic data demonstrates its capabilities and potential impact. In many circumstances where healthcare solutions are being developed, real patient data in not available to developers. Thus, we generated synthetic Irish health chronic disease datasets based on the metadata from HIPE, CDM, PCRS, and Restina in to use in our validation studies. As part of this validation, 14 new healthcare assets were created to illustrate how clinicians, strategists and policy makers can benefit from the deployment of the RECONNECT prototype.

\section*{Acknowledgement}
The research is part of the RECONNECT project: chRonic disEase: disCOvery, aNalysis aNd prEdiCTive modelling. This publication has emanated from research conducted with the financial support of Science Foundation Ireland under Grant number [22/NCF/DR/11244 and 12/RC/2289\_P2]. For the purpose of Open Access, the author has applied a CC BY public copyright licence to any Author Accepted Manuscript version arising from this submission


\end{document}